\documentclass[12pt]{article}
\usepackage{epsfig,amsfonts,amssymb}
\usepackage{hyperref}
\newcommand{\be}{\begin{equation}}
\newcommand{\ee}{\end{equation}}
\newcommand{\ben}{\begin{eqnarray}}
\newcommand{\een}{\end{eqnarray}}
\newcommand{\ba}{\begin{eqnarray}}
\newcommand{\ea}{\end{eqnarray}}
\newcommand{\nn}{\nonumber \\}

\newcommand{\beq}{\begin{equation}}
\newcommand{\eeq}{\end{equation}}

\newcommand{\ie}{{\it i.e.,}\ }

\newcommand{\mt}[1]{\textrm{\tiny #1}}

\newcommand{\seff}{S_\mt{eff}}
\newcommand{\bi}{\begin{itemize}}
\newcommand{\ei}{\end{itemize}}
\newcommand{\ii}{\item}

\newcommand{\rhic}{{\bf RHIC} }
\newcommand{\adscft}{{\bf AdS/CFT}}
\newcommand{\ads}{{\bf AdS}}
\newcommand{\qcd}{{\bf QCD}}
\newcommand{\etabys}{${\eta \over s}$}
\newcommand{\kss}{{\bf KSS}}
\newcommand{\rd}{re-definition}
\newcommand{\apt}{$\alpha'$}
\newcommand{\aps}{$\alpha'^{2}$}
\newcommand{\rims}{${\bf Riemann^{2}}$}
\newcommand{\rics}{${\bf Ricci^{2}}$}
\newcommand{\rs}{${\bf R^{2}}$}
\newcommand{\ff}{(g_{(2)} + h_{(2)})}

\newcommand{\lb}{\left (}
\newcommand{\rb}{\right )}
\newcommand{\ltb}{\left [}
\newcommand{\rtb}{\right ]}
\newcommand{\ep}{\epsilon}

\newcommand{\cA}{{\cal A}}
\newcommand{\cB}{{\cal B}}
\newcommand{\ph}{\phi(r,k)}
\newcommand{\php}{\phi'(r,k)}
\newcommand{\cph}{\phi(r,-k)}
\newcommand{\cphp}{\phi'(r,-k)}
\newcommand{\phpp}{\phi''(r,k)}
\newcommand{\cphpp}{\phi''(r,-k)}
\newcommand{\p}{\partial}
\newcommand{\ra}{\rightarrow}
\newcommand{\intk}{\int {d^4 k \over (2 \pi)^4}}
\newcommand{\nt}{{1\over 16 \pi G_5}}

\newcommand{\bc}{\begin{center}}
\newcommand{\ec}{\end{center}}

\newcommand{\hpt}{\hspace{2cm}}

\newcommand{\ebs}{${\eta \over s}$\ }
\newcommand{\ap}{\alpha'}
\newcommand{\sectiono}[1]{\section{#1}\setcounter{equation}{0}}

\newcommand{\gz}{g_{(0)}}
\newcommand{\go}{g_{(1)}}
\newcommand{\gt}{g_{(2)}}
\newcommand{\tr}{{\rm Tr}}
\newcommand{\cu}{\alpha'}

\def\one{{\hbox{ 1\kern-.8mm l}}}
\def\zero{{\hbox{ 0\kern-1.5mm 0}}}

\begin{document}

\baselineskip 24pt

\begin{center}
{\Large \bf Shear Viscosity to Entropy Density Ratio in Six Derivative Gravity
}

\end{center}

\vskip .6cm
\medskip

\vspace*{4.0ex}

\baselineskip=18pt

\centerline{\large \rm   $^{a}$Nabamita Banerjee, $^{b}$Suvankar Dutta}

\vspace*{4.0ex}

\centerline{\large \it $^{a}$ Harish-Chandra Research Institute }

\centerline{\large \it  Chhatnag Road, Jhusi,
Allahabad 211019, INDIA}
\vspace{.5cm}
\centerline{\large \it $^{b}$ Dept. of Mathematical and Statistical
Science,}

\centerline{\large \it University of Alberta,
  Canada}

\vspace*{1.0ex}
\centerline{E-mail:  $^{a}$ nabamita at hri.res.in, $^{b}$
  sdutta at math.ualberta.ca}

\vspace*{5.0ex}

\centerline{\bf Abstract} \bigskip

We calculate shear viscosity to entropy density ratio in presence of
four derivative (with coefficient $\alpha'$) and six derivative (with
coefficient $\alpha'^2$) terms in bulk action. In general, there can
be three possible four derivative terms and ten possible six
derivative terms in the Lagrangian. Among them two four derivative and
eight six derivative terms are ambiguous, i.e., these terms can be
removed from the action by suitable field redefinitions. Rest are
unambiguous. According to the AdS/CFT correspondence all the
unambiguous coefficients (coefficients of unambiguous terms) can be
fixed in terms of field theory parameters. Therefore, any measurable
quantities of boundary theory, for example shear viscosity to entropy
density ratio, when calculated holographically can be expressed in
terms of unambiguous coefficients in the bulk theory (or equivalently
in terms of boundary parameters). We calculate $\eta/s$ for generic six derivative gravity and find that apparently it depends on few ambiguous coefficients at order $\alpha'^2$.
We calculate six derivative corrections to central charges $a$ and $c$ and express $\eta/s$ in terms of these central charges and unambiguous coefficients in the bulk theory.

\newpage

\tableofcontents


\sectiono{Introduction and Summary}

One of the current interests, in the context of \adscft, is to investigate different properties of quark-gluon plasma
({\bf QGP}) created at the Relativistic Heavy Ion Collider ({\bf RHIC}).
The temperature of the gas of quarks and gluons produced at \rhic \ is approximately $170 MeV$ which is very close to the confinement temperature of \qcd. Therefore, at this high temperature they are not in the weakly coupled regime of \qcd. In fact near the transition temperature the gas of quarks and gluons belongs to the non-perturbative realm of \qcd, where one can not apply the result of perturbative QFT to study their properties. Different kinetic
coefficients of this strongly coupled plasma is not possible to calculate with the usual set up of perturbative \qcd. The \adscft \ correspondence \cite{malda,gkp,witten}, at this point, appears as a technically powerful tool to deal with strongly coupled ($\it conformal$) field theory in terms of weakly coupled (super)-gravity theory in {\bf AdS} space. The \adscft \ can be an approximate representation of \qcd \ only at high enough temperature since \qcd \ does not have any conformal invariance ($\beta$ function is not $zero$). However, we assume that the \qcd \ plasma is well described by some conformal field theory which has a gravity dual.

The first success in this direction came from the holographic calculation of shear viscosity coefficient of conformal gauge theory plasma in the context of \adscft \ \cite{pss}. Other transport coefficients of dual gauge theory have also been calculated in the \adscft \ framework \cite{visco1}-\cite{viscoend}. In this paper we will concentrate on an interesting conformally invariant measurable parameter of gauge theory plasma, namely, shear viscosity to entropy density ratio (${\eta \over s}$). The primary motivation for this particular ratio is following. A large class of gauge theories with gravity dual have ${\eta \over s}= {1\over 4 \pi}$ which is in a good agreement with \rhic \ data.

In \cite{kss} Kovtun, Son and Starinets have conjectured that the ratio \etabys \ has a lower bound (\kss\ bound)
\be
{\eta \over s} \geq {1\over 4 \pi}
\ee
for all relativistic quantum field theories at finite temperature and the inequality is saturated by theories with gravity dual \ie without any higher derivative corrections. The leading $\alpha'$ correction coming from type II string theory is $R^4$ term. In has been shown in \cite{buchel1,buchel2} that the presence of $R^4$ term in the action increases the value of \etabys \ beyond ${1\over 4 \pi}$. But the story is different when one considers four derivative terms in the bulk action. These terms appear in Heterotic string theory. It has been shown in \cite{kats,myers1} that four derivative terms actually decreases the value of \etabys \ bellow the lower bound. In \cite{kats}, authors proposed an example of string theory model where the conjectured bound is violated.

An explicit and more detailed investigation on violation of \kss\ bound has been studied in \cite{myers-sinha} in the context of four derivative gravity. A generic four derivative action can have three terms : \rims, \rics\ and \rs ({\bf R} is Ricci scalar). Second and third term can be removed by field \rd. Therefore we are left with two independent parameters: coefficients of \rims \ and (dimension less) radius of \ads \ space. \cite{myers-sinha} found relations between these two parameters in gravity side and two parameters in the boundary theory, namely the central charges $c$ and $a$. Hence \etabys \ can be expressed in terms of these two central charges. Therefore they argued that the violation of \kss \ bound depends on these two central charges of boundary conformal field theory. First of all the central charges should satisfy two conditions: $c\sim a \gg 1$ and $|c-a|/c\ll 1$ and then the bound is violated when $c-a > 0$.

Though it is possible to determine these two parameters in the bulk action and hence \etabys \ in terms of two central
charges of boundary theory in four derivative case but in a generic higher derivative gravity it is not obvious how to express \etabys \ in terms of independent boundary parameters. For example, in this paper we consider generic six derivative terms in bulk. These six derivative terms do not appear in any super-string (type IIA or IIB) or heterotic theory but they can arise in bosonic string theory \cite{loopcorrection}. Therefore it is quite interesting to study the effects of these terms on the hydrodynamic behavior of boundary gauge theory plasma, in particular on the ratio \etabys. Needless to mention, the gauge theory plasma is not super-symmetric in this case. There can be total ten possible six derivative terms with different coefficients in bulk Lagrangian.
We call those coefficients (or terms) {\bf "ambiguous"} which can be removed from the effective action by some field
\rd\ and other coefficients (or terms) which can not be removed by any field \rd\ we refer them {\bf "unambiguous"}.
It is possible to show that among ten different terms eight of them can be removed after a suitable field
re-definition \cite{mt}. Therefore the bulk theory has two unambiguous (six derivative) coefficients (we denote them by $\alpha_1$ and $\alpha_2$). If we assume that the effective bulk theory has a dual field theory description then different parameters of boundary conformal field theory, which capture its aggregate properties, should be able to fix the unambiguous couplings of dual gravity theory. In other words, all the unambiguous coefficients of bulk theory can
be expressed in terms of physical boundary parameters. For example in
\cite{hofmal} authors found that a combination of
$\alpha_1$ and $\alpha_2$ (namely $2 \alpha_1 + \alpha_2$) is related to a coefficient (we denote it by $\tau_4$) in field theory which appears in correlation
of energy one point function (three point function of energy momentum
tensor).  We discussed about this in brief details in section [\ref{etasandetabys}].
Therefore any measurable quantities of boundary theory, for example shear viscosity to entropy density ratio, when calculated holographically should be expressed in terms of unambiguous coefficients in the bulk theory or boundary parameters.

We  calculate the ratio \etabys\ for generic six derivative terms.
It turns out that the ratio depends on two ambiguous coefficients (we call them $\alpha_3$ and $\alpha_4$). In section [\ref{redefsec}] we have discussed these in details. The $apparent$ dependence on ambiguous coefficients in physical quantities is an artifact of our choice of starting Lagrangian. One could start with a Lagrangian where all the ambiguous coefficients are set to zero. In that case, shear viscosity coefficient, entropy density and their ratio would be independent of these ambiguous coefficients. However, for being more explicit we start with the most generic Lagrangian and find that the physical quantities like $\eta$, $s$ and \etabys\ depend on some ambiguous coefficients.
 Therefore it seems to be puzzling how to express these quantities completely in terms of boundary parameters. In this paper we show that it is still possible to express $\eta$, $s$ and \etabys\ in terms two central charges $a$ and $c$ and other two unambiguous coefficients\footnote{We assume that the "unambiguous" coefficients of higher derivative gravity can be fixed by boundary parameters.} $\alpha_1$ and $\alpha_2$.  Our final results are\footnote{We are thanful to R. Myers for pointing out a mistake in Eq. (\ref{etaca}) and Eq. (\ref{sca}) in the previous version.}
\ben \label{etaca}
\eta &=& 8 \pi^3 c \ T^3 \bigg[ 1 + {1\over 4}{c-a \over c} - {1\over 8} \bigg( {c-a \over c}\bigg)^2- {180 \over \lambda} (2 \alpha_1 +  \alpha_2)\bigg]  + {\cal O}(\lambda^{-3/2})\ ,
\een
\ben\label{sca}
s &=& 32 \pi^4 c\  T^3 \bigg[ 1 + {5\over 4} {c-a \over c}  +
 {3 \over
8}\bigg({c-a \over c}\bigg)^2+ {12 \over \lambda} (2 \alpha_1+ \alpha_2)\bigg]  + {\cal O}(\lambda^{-3/2})
\een
 and
\be \label{finalres}
{\eta \over s} = {1\over 4 \pi}\bigg[1 -{c-a \over  c} + {3\over 4} \bigg({c-a \over c}\bigg)^2 - {192\over \lambda} (2 \alpha_1 + \alpha_2) \bigg] + {\cal O}(\lambda^{-3/2})\ ,
\ee
where $T$ is the temperature and $\lambda$ is the 't Hooft coupling.

We obtain this result in the following way. Since six derivative terms appear with coefficient \aps \ where four derivative terms are proportional to \apt,
therefore to make all the expressions correct up to order \aps, we need to consider the effect of four derivative terms to order \aps \ also. As we mentioned earlier at order \apt, the coefficients of \rs\ and \rics\ terms ($\beta_1$ and $\beta_3$ respectively) are ambiguous, they can be removed by field \rd\ \cite{mt}. In fact they do not appear in the expression of \etabys \ at order \apt. But these two ambiguous coefficients appear at order \aps\ (see section [\ref{redefsec}]). Therefore the ratio \etabys\ depends on three unambiguous coefficients $\beta_2$ (at order \apt), $\alpha_1$ and  $\alpha_2$ (at order \aps) and four ambiguous coefficients $\beta_1, \beta_3, \alpha_3$ and $\alpha_4$ at order \aps.  Then we calculate two central charges $a$ and $c$ for six derivative gravity.
We consider a particular combination of these central charges, namely $c-a \over c$. It turns out that the combinations of ambiguous coefficients, which appear in the expression of \etabys, the same combination appears in $c-a \over c$. Therefore one can remove all ambiguous coefficients in terms of this particular combination of central charges $a$ and $c$.

Let us summarize the main results of this paper.
\bi
\ii
In section [\ref{redefsec}] we consider the most general six derivative action. There can be total ten independent invariants. We identify the ambiguous and unambiguous coefficients of this generic action. We find that it is possible to drop six ambiguous terms from the action on which \etabys\ does not depend. We also consider the effect of four derivative terms to order \aps.
\ii
In section [\ref{pertmet}] we calculate the perturbed background metric up to order \aps.
\ii
In section [\ref{effacn}] we compute the ratio \etabys\ using effective action approach of \cite{bd}.
\ii
In section [\ref{ca}] we calculate the central charges $a$ and $c$ for six derivative gravity.
\ii
Finally in section [\ref{etasandetabys}] we write the expression for $\eta$, $s$ and \etabys\ in terms of central charges and two unambiguous parameters of bulk Lagrangian. We also discuss how to relate the unambiguous coefficients of bulk theory to the physical boundary parameters following \cite{hofmal}.
\ii
In appendix [\ref{As}] and [\ref{cb0cb1}] we present the expressions for $A_i$'s and $\cB$'s respectively which appear in section [\ref{effacn}].
\ii
We also calculate shear viscosity coefficient using Kubo formula as a check of our effective action calculation. In appendix [\ref{kubo}] we outline the calculations.
\ii
In appendix [\ref{apprdep}] we calculate leading $r$ dependence of Riemann and Ricci tensors which appear in section [\ref{ca}].
\ei


\sectiono{The Field Re-definition and ${\eta \over s}$}
\label{redefsec}

In this section we discuss the most general six derivative terms in the bulk Lagrangian and their effects on shear viscosity to entropy density ratio. Generic six derivative terms can be constructed out of Riemann tensor, Ricci tensors and curvature scalar terms or their covariant derivatives.
There are five possible dimension-6 invariants which do not involve Ricci tensors or curvature scalars,
\ben
I_1 &=&
R^{\mu\nu}_{\, \, \, \, \alpha\beta}R^{\alpha\beta}_{\, \, \, \,
  \lambda\rho}R^{\lambda\rho}_{\, \, \, \, \mu\nu},
\nonumber \\
I_2&=&
R^{\mu\nu}_{\, \,\, \, \rho\sigma}R^{\rho\tau}_{\, \, \, \, \lambda\mu} R^{\sigma \ \ \lambda}_{\ \ \tau \ \ \nu},
\nonumber \\
I_3&=&
R^{\alpha\nu}_{\, \,\, \, \mu\beta}R^{\beta\gamma}_{\, \, \, \, \nu\lambda}
R^{\lambda\mu}_{\, \,\,\, \gamma\alpha},
\nonumber \\
I_4&=&R_{\mu\nu\alpha\beta}R^{\mu \alpha}_{\, \, \, \, \gamma\delta}
R^{\nu\beta\gamma\delta},
\nonumber \\
I_5&=&R_{\mu\nu\alpha\beta}{\cal D}^2 R^{\mu\nu\alpha\beta}\ .
\een
They satisfy the following relations,
\be
I_3= I_2-{1\over 4}I_1, \quad I_4={1\over 2}I_1, \quad I_5 = -I_1 - 4 I_2 \ .
\ee
Hence only two of them are independent.  We will choose these two invariants to be $I_1$
and $I_2$.

Now consider the most general action containing all possible independent
curvature invariants
\ben \label{6dacn}
{\cal I} &=& \int d^5x \sqrt{-g} \ {\cal L}
\een
where
\ben \label{6dlag}
{\cal L} &=& a_0 R - 2 \Lambda +
 \alpha'  \bigg (  \beta_1 R^2 + \beta_2
R_{\mu\nu\rho\sigma} R^{\mu\nu\rho\sigma} + \beta_3 R_{\mu\nu}R^{\mu\nu}  \bigg ) \nonumber \\
&& + \alpha'^2 \bigg ( \alpha_1 I_1 + \alpha_2 I_2 + \alpha_3
R_{\mu\nu\beta\gamma}R^{\beta\gamma\nu\rho}R^{\mu}_{\rho} + \alpha_4
R R_{\mu\nu\rho\sigma}R^{\mu\nu\rho\sigma} + \alpha_5
R_{\mu\nu\rho\lambda} R^{\nu\lambda}R^{\mu \rho}
\nonumber \\
&& + \alpha_6 R_{\mu \nu} R^{\nu
  \lambda} R^{\mu}_{\, \, \lambda}+ \alpha_7 R_{\mu\nu}{\cal D}^2 R^{\mu\nu}  + \alpha_8 R R_{\mu\nu}R^{\mu\nu} + \alpha_9
R^3 + \alpha_{10} R {\cal D}^2 R \bigg ) + {\cal O}(\alpha'^3)\ . \nn
\een
However, this action is ambiguous up to a field re-definition. It has been
shown in \cite{mt}
that under the following field re-definition
\ben \label{redef1}
g_{\mu \nu} \ra {\tilde g}_{\mu\nu} &=& g_{\mu \nu}
+ \alpha' \big (d_1 g_{\mu\nu} R + d_2 R_{\mu\nu}
\big )
\nonumber \\
&& \quad\ \   +
\alpha'^2 \big (d_3 R_{\mu \alpha \beta \gamma}R_{\nu}^{\, \, \alpha \beta
\gamma} + d_4 g_{\mu\nu}
R_{\alpha\beta\gamma\sigma} R^{\alpha\beta\gamma\sigma}+
d_5 R_{\mu \alpha\beta\nu}R^{\alpha\beta}  + d_6 R_{\mu \lambda}R^{\lambda}_{\,
\, \nu} \nonumber \\
&& \quad \ \  + d_7 {\cal D}^2 R_{\mu\nu}
 + d_8 g_{\mu\nu}
R_{\alpha\beta}R^{\alpha\beta} + d_9 g_{\mu\nu} R^2 + d_{10} g_{\mu\nu} {\cal
  D}^2 R \big ) + {\cal O}(\alpha'^3)\nonumber \\
\een
the coefficients $a_0,\beta_2, \alpha_1$ and $\alpha_2$ in the Lagrangian (\ref{6dlag}) remain invariant and all other
coefficients changes. This is because it is not possible to generate any higher rank tensor from a lower rank tensor in (\ref{redef1}). For example one can not get ${\bf Riemann^2}$ term from any ${\bf Ricci}$ term at order $\alpha'$ and similarly any ${\bf Riemann^3}$ term can not be generated from any ${\bf Ricci^2}$, ${\bf Rieman^2}$ or ${\bf Ricci \cdot Riemann}$ terms at order $\alpha'^2$. Therefore the coefficients $\beta_2, \alpha_1$ and $\alpha_2$ are unambiguous.
By proper choice of $d_1, ..., d_{10}$ one can set any
desired values to the coefficients $\beta_1, \beta_3$ and $\alpha_3, ..., \alpha_{10}$, for example we can set all of them to zero. These are the ambiguous coefficients. Setting all ambiguous coefficients to zero the action (\ref{6dacn}) becomes,
\ben \label{6dacnspe}
\sqrt{-g}\ {\cal L} &\ra&  \sqrt{-g} \bigg(\tilde a_0 R - 2 \Lambda +
 \alpha'  \ \beta_2
R_{\mu\nu\rho\sigma} R^{\mu\nu\rho\sigma}
\ + \alpha'^2 \bigg ( \alpha_1 I_1 + \alpha_2 I_2 \bigg)\bigg)
\een
with some different $\tilde a_0$ which is related to $a_0$ and other
ambiguous parameters\footnote{$\tilde a_0$ gets contribution from $\sqrt{-g}$.}. The action (\ref{6dacnspe}) and (\ref{6dacn}) are equivalent up to a field \rd. Any physical quantity like entropy, shear viscosity or their ratio calculated either from action (\ref{6dacnspe}) or (\ref{6dacn}) turns out to be same after using the relation between $a_0$ and $\tilde a_0$. That is, these quantities are field \rd\ invariant.

We calculate \etabys\ for generic six derivative action and find that the ratio depends on some ambiguous coefficients in (\ref{6dacn}). Before we start calculating \etabys\ for the generic action (\ref{6dacn}) we can use the following logic to understand that among ten ambiguous parameters six of them never appear in the expression of \etabys\footnote{Though these coefficients may arise in the individual expressions of $\eta$ and $s$. Since we are interested in \etabys\ we drop these terms. However the final expressions (\ref{etaca} and \ref{sca}) for $\eta$ and $s$ remain unchanged even if we consider these terms.}. Therefore we can drop those terms at the beginning to simplify our life\footnote{Other ambiguous terms can not be dropped using this logic.}. Let us now find out those terms in the action on which \ebs does not depend.

Consider the following Lagrangian,
\ben \label{acnnotreq}
{\tilde {\cal L}} &=& a_0 R - 2 \Lambda  + \alpha'^2 \bigg (\alpha_5
R_{\mu\nu\rho\lambda} R^{\nu\lambda}R^{\mu \rho}
+ \alpha_6 R_{\mu \nu} R^{\nu
  \lambda} R^{\mu}_{\, \, \lambda}+ \alpha_7 R_{\mu\nu}{\cal D}^2 R^{\mu\nu}  \nonumber \\
&& \hpt \qquad \ \ + \alpha_8 R R_{\mu\nu}R^{\mu\nu} + \alpha_9
R^3 + \alpha_{10} R {\cal D}^2 R \bigg ) + {\cal O}(\alpha'^3)
\een
and following field re-definition,
\ben \label{redef2}
g_{\mu \nu} &\ra&  g_{\mu \nu}
+ \alpha'^2 \big (d_5 R_{\mu \alpha\beta\nu}R^{\alpha\beta}  + d_6 R_{\mu \lambda}R^{\lambda}_{\,
\, \nu} \nonumber \\
&& \quad \ \  + d_7 {\cal D}^2 R_{\mu\nu}
 + d_8 g_{\mu\nu}
R_{\alpha\beta}R^{\alpha\beta} + d_9 g_{\mu\nu} R^2 + d_{10} g_{\mu\nu} {\cal
  D}^2 R \big ) + {\cal O}(\alpha'^3)\ .
\een
With proper choice of $d_5, d_6$...$d_{10}$ one can check that the resultant Lagrangian becomes,
\be \label{acn0}
\sqrt{-g}{\tilde {\cal L}} \ra \sqrt{-g}({\tilde a}_0 R - 2 \Lambda) \ .
\ee
Also under the field re-definition (\ref{redef2}) the metric scales in the following way,
\be \label{metscl}
g_{\mu\nu} \ra {\cal C} (\alpha') g_{\mu\nu}
\ee
where,
\be
{\cal C} (\alpha')= 1 + \alpha'^2 (-16 d_5 + 16 d_6 + 80 d_8 + 400 d_9)\ .
\ee
Here we have used the leading equation of motion $R_{\mu\nu}=-4 g_{\mu\nu}$.
The scaling in (\ref{metscl}) does not change the temperature of the background spacetime and hence the diffusion pole calculated from action (\ref{acn0}) gives the standard result $D={1\over 4 \pi T}$, where $D$ is diffusion constant and $T$ is temperature. Thus the ratio \ebs turns out to be ${1\over 4 \pi}$ for action (\ref{acnnotreq}). Therefore we see that shear viscosity to entropy density ratio does not depend on $\alpha_5, \alpha_6 \cdot \cdot \cdot \alpha_{10}$ up to order $\alpha'^2$.

One important thing to notice here is that the ratio \ebs does not depend on $\beta_1$ and $\beta_3$ up to order $\alpha'$ \cite{myers1,kats}. One can consider the following field re-definition
\be \label{metscalegb}
g_{\mu\nu} \ra g_{\mu\nu} + \alpha' (d_1 g_{\mu\nu} R + d_2 R_{\mu\nu})
\ee
and get rid off the terms $\beta_1 R^2$ and $\beta_3 R_{\mu\nu}^2$ with proper choice of $d_1$ and $d_2$. The new metric is same as the original metric up to some constant scaling factor {\it to order} $\alpha'$ (substituting the leading equation of motion at order \apt). Therefore one can argue that \etabys\ is independent of $\beta_1$ and $\beta_3$ up to order \apt. But this is not true when we consider terms to order $\alpha'^2$. We can not only substitute the leading order equation of motion in (\ref{metscalegb}) when we are interested in \aps\ order. We have to consider equations of motion to order $\alpha'$. The equations of motion to order $\alpha'$ is given by \cite{nd},
\ben
R_{\mu\nu} &=& -4 g_{\mu\nu} + {\alpha'\over 3}  L^{(2)} g_{\mu\nu} -2 \alpha' L^{(2)}_{\mu\nu} -\ap ( \beta_3 + 4 \beta_2) {\cal D}^2 R_{\mu\nu} \nonumber \\
&& \ \ + {2 \ap \over 3} (3\beta_1 + \beta_3 + \beta_2) g_{\mu\nu}  {\cal D}^2 R + \ap (2 \beta_1 + \beta_3 + 2 \beta_2) D_{\mu\nu}R + {\cal O}(\alpha'^2)\ .
\een
Substituting this equations of motion in (\ref{metscalegb}) we get,
\ben
g_{\mu\nu} &\ra& g_{\mu\nu} -4  \ap (5 d_1 + d_2) g_{\mu\nu}   \nonumber \\
&& \ \ \ + \ap^2 \ltb {d_2 -d_1 \over 3} (400 \beta_1 + 80 \beta_3)- 2 d_2 (16 (2 \beta_2 + \beta_3) - 32 \beta_2 + 80 \beta_1) \rtb g_{\mu\nu}\nonumber \\
&& \ \ \ + \ap^2 \beta_2 \ltb  {d_2 -d_1 \over 3} R_{\mu\nu\rho\sigma}^2 g_{\mu\nu}- 2 d_2 R_{\mu\alpha\beta\gamma}R_{\nu}^{\ \ \alpha\beta\gamma} \rtb \ .
\een
Therefore we see that the new metric is proportional to the original metric (with constant proportionality factor) at order $\alpha'$ but not at order $\alpha'^2$ when $\beta_2 \neq 0$. Hence \etabys \ may not be independent of $\beta_1$ and $\beta_3$ at order \aps. It can have terms like $\beta_1 \beta_2$, $\beta_2 \beta_3$ and $\beta_2^2$ at order \aps.


\sectiono{The Perturbed Background Metric} \label{pertmet}

In this section we will find the perturbative solution to Einstein equations in presence of six derivative terms in the action up to order \aps. We write the basic equation of motions and mention how to solve these equations up to order \aps.
We will start with the following five dimensional action with negative cosmological constant $\Lambda=-6$.
\ben \label{6dacnreqg}
{\cal I} &=& \nt \int d^5x \sqrt{-g} \bigg [ R - 2 \Lambda +
 \ap  \bigg (  \beta_1 R^2 + \beta_2
R_{\mu\nu\rho\sigma} R^{\mu\nu\rho\sigma} + \beta_3 R_{\mu\nu}R^{\mu\nu}  \bigg ) \nonumber \\
&& + \ap^2 \bigg ( \alpha_1 I_1 + \alpha_2 I_2 + \alpha_3
R_{\mu\alpha\beta\gamma}R^{\beta\gamma\alpha\rho}R^{\mu}_{\rho} + \alpha_4
R R_{\mu\nu\rho\sigma}R^{\mu\nu\rho\sigma} \bigg) \bigg]\ .
\een
We take the leading value of of \ads \ radius is 1.

We consider the following metric ansatz (assuming planer symmetry of the spacetime),
\be
ds^2= -\rho^2 \mathrm{e}^{2A(\rho) + 8 B(\rho)} dt^2 + \rho^2 \mathrm{e}^{2 B(\rho)} d\rho^2 + \rho^2 d\vec{x}^2\ .
\ee
Substituting this metric in the (\ref{6dacnreqg}) we get,
\be
{\cal I} = \nt \int_{\rho_0}^{\infty} d\rho \ltb \mathfrak{L}^{(2)} + \ap \mathfrak{L}^{(4)} + \ap^2 \mathfrak{L}^{(6)} \rtb
\ee
where,
\ben
\mathfrak{L}^{(2)} &=& \sqrt{-g} (R + 12) \nonumber \\
&=& 12 \rho^5 \mathrm{e}^{A(\rho) + 5 B(\rho)} - 2 \rho (2 + \rho A'(\rho)) \mathrm{e}^{A(\rho) + 3 B(\rho)} - 2 {d\over d\rho} \ltb (A'(\rho) + 4 B'(\rho)) \rho^3 \mathrm{e}^{A(\rho) + 3 B(\rho)}\rtb\nonumber \\
\een
and $\mathfrak{L}^{(4)}$ and $\mathfrak{L}^{(6)}$ are four and six derivative terms in the Lagrangian evaluated on the metric ansatz.
The Euler-Lagrange equations which follow from this action is given by,
\ben \label{ELeom}
{\p \mathfrak{L}^{(2)} \over \p A(\rho)} -{d\over d\rho}{\p \mathfrak{L}^{(2)} \over \p A'(\rho)} &=& -\ap \lb {\p \mathfrak{L}^{(4)} \over \p A(\rho)} - {d\over d\rho}{\p \mathfrak{L}^{(4)} \over \p A'(\rho)} + {d^2\over d\rho^2}{\p \mathfrak{L}^{(4)} \over \p A''(\rho)} \rb\nonumber \\
&&-\ap^2 \lb {\p \mathfrak{L}^{(6)} \over \p A(\rho)} - {d\over d\rho}{\p \mathfrak{L}^{(6)} \over \p A'(\rho)} + {d^2\over d\rho^2}{\p \mathfrak{L}^{(6)} \over \p A''(\rho)} \rb\nonumber \\
{\p \mathfrak{L}^{(2)} \over \p B(\rho)} &=& -\ap \lb {\p \mathfrak{L}^{(4)} \over \p B(\rho)} - {d\over d\rho}{\p \mathfrak{L}^{(4)} \over \p B'(\rho)} + {d^2\over d\rho^2}{\p \mathfrak{L}^{(4)} \over \p B''(\rho)} \rb\nonumber \\
&&-\ap^2 \lb {\p \mathfrak{L}^{(6)} \over \p B(\rho)} - {d\over d\rho}{\p \mathfrak{L}^{(6)} \over \p B'(\rho)} + {d^2\over d\rho^2}{\p \mathfrak{L}^{(6)} \over \p B''(\rho)}\rb\ .
\een
We solve this equation perturbatively to find $A(\rho)$ and $B(\rho)$ . First we solve this equations up to order \apt. We use leading order solutions for $A$ and $B$ on the right hand side. The order $\ap$ terms on the right hand side will act as a source terms and we solve the equations to find corrected $A$ and $B$ in presence of these source terms. There are two integration constants when we solve this equations. We choose these two integration constants (to order $\ap$) such a way that the corrected (black hole)solution has horizon at $\rho=1$ and the boundary ($\rho \ra \infty$) metric
is Minkowskian.

After getting the metric up to order $\ap$ we now solve $A$ and $B$ to order $\ap^2$. We substitute the solutions for $A$ and $B$ (corrected up to order $\ap$) on the right hand side of equation (\ref{ELeom}) and get the solution for $A$ and $B$ to order $\ap^2$. We again choose the integration constants in order to set the black hole horizon radius at $\rho=1$ and the boundary metric to be Minkowskian.

The solution is given by (after changing the coordinate $\rho\ra {1 \over \sqrt{r}}$),
\be \label{6derimet}
ds^2 =  f(r) dt^2 + {g(r)\over 4 r^3} dr^2 + {1\over r} d\vec{x}^2
\ee
where $f(r)$ and $g(r)$ are given by,
\ben
f(r)&=&r - \frac{1}{r} -
 2 r\big (r^2 - 1 \big)\beta_2 \ap \nonumber \\
 && +\frac {1} {3} r\big (r^2 -
     1 \big)\bigg (12 (2r^2 -31) \alpha_1 + (48r^2 -33) \alpha_2 + 24 (2r^2 +3) \alpha_3 - 24 (12 r^2 + 7) \alpha_4 \nn
     && + 4 \beta_2(-22 \beta_1 + 48 r^2 \beta_1 + 149 \beta_2 - 42 r^2 \beta_2 + 34 \beta_3 )\bigg ) \ap^2
\een
and
\ben
g(r) &=&  \frac {r} {1- r^2} + \frac {2 r \big (
      10 \beta_1 + (1- 3 r^2) \beta_2 +
      2 \beta_3 \big) \ap} { 3(r^2 - 1)}  \nonumber \\
      && + {r \over 9 (r^2 -
      1 )} \bigg ( 12 (1 - 93 r^2 + 240 r^4) \alpha_1 + 9 (1 - 11r^2 - 2 r^4) \alpha_2 \nn
      && - 24 (1 - 9 r^2 + 36 r^4) \alpha_3 + 24 (5 - 21 r^2 + 126 r^4 ) \alpha_4 + 400 \beta_1^2 \nn
      && + 16  (5 - 9 r^2 - 126 r^4) \beta_1 \beta_2
      + 4( 1 + 450 r^2 - 927 r^4) \beta_2^2 + 160 \beta_1 \beta_3 \nn
      && + 16 (1 + 27 r^2 - 90 r^4) \beta_2 \beta_3 + 16 \beta_3^2 \bigg ) \ap^2\ .
\een
This is the background metric corrected up to order \aps. Also the black brane temperature is given by,
\ben\label{temp}
T&=& {1 \over \pi} +{ 10 \beta_1-5 \beta_2+ 2 \beta_3 \over 3 \pi} \cu\nn
&& +{1 \over 18 \pi } \bigg (732 \alpha_1 -63 \alpha_2 -312 \alpha_3 +1272 \alpha_4 + 700
  \beta_1^2 \nn
  && -1948 \beta_1 \beta_2 -605 \beta_2^2 + 280 \beta_1 \beta_3
-620 \beta_2 \beta_3 + 28 \beta_3^2 \bigg) \cu^2 \ .
\een


\sectiono{The Effective Action and Shear Viscosity}\label{effacn}

To calculate six derivative correction to the shear viscosity coefficient we need to find the quadratic action for transverse graviton moving in background spacetime (\ref{6derimet}). We consider the following metric perturbation,
\be\label{petmet}
g_{xy}=g^{(0)}_{xy}+ h_{xy}(r,x)=g^{(0)}_{xy}(1+\ep \Phi(r,x))
\ee
where $\epsilon$ is an order counting parameter. We  consider terms up to
order $\epsilon^2$ in the action of $\Phi(r,x)$.
The action (in momentum space) is given by,
\ben \label{gravacn}
S&=&\nt \int {d^4 k \over (2 \pi)^4} dr \bigg [A_1(r,k)
\ph \cph + A_2(r,k) \php
\cphp \nonumber \\
&& \hspace {2cm}
+ A_3(r,k) \phpp \cphpp + A_4(r,k) \ph \cphp \nonumber \\
&& \hspace{2cm}+ A_5(r,k) \ph \cphpp
+ A_6(r,k)
\php \cphpp \bigg ]
\een
where the expressions for $A_i$s are given in appendix [\ref{As}] and $\ph$ is given by,
\be \label{phifu}
\ph = \int {d^4x \over (2 \pi)^4} e^{-i k.x} \Phi(r,x)\ ,
\ee
$k=\{-\omega,{\vec
k}\}$ and `\ $'$\ ' denotes derivative with respect to $r$.
Up to some total
derivative terms this action can be written as,
\ben \label{gravacng}
S= \nt \int {d^4 k \over (2 \pi)^4} dr \bigg [ \cA_0 \ph \cph + \cA_1
\php \cphp + \cA_2 \phpp \cphpp \bigg ]
\een
where,
\ben
\cA_0 &=& A_1(r,k) -{A'_4(r,k) \over 2} + {A_5''(r,k)
\over 2}
\nonumber \\
\cA_1 &=& A_2(r,k) - A_5(r,k) -{A'_6(r,k) \over 2}
\nonumber \\
\cA_2 &=& A_3(r,k) \ .
\een

This action does not have the canonical form. Therefore to obtain the shear viscosity coefficients from this action we follow the prescription given in \cite{bd}. We write the effective action for the scalar field,
\ben \label{gravacneff}
\seff &=& {1 \over 16 \pi G_5} \intk \bigg [
 (\cA^{(0)}_1(r,k) +
\ap \cB_1^{(0)}(r,k) + \ap^2 \cB_1^{(1)}(r,k)) \cphp \php \\
&& \hspace{2cm}
+ (\cA^{(0)}_0(r,k)+\ap \cB_0^{(0)}(r,k) + \ap^2 \cB_0^{(1)}(r,k)) \ph \cph \bigg ]\ .
\een
where,
\be
\cA_1^{(0)}(r) = {r^2 -1 \over r}
\ee
and
\be
\cA_0^{(0)}(r,k)= {\omega^2 \over 4r^2(1-r^2)}\ .
\ee
To evaluate the functions $\cB_1^{(0)}, \cB_1^{(1)}, \cB_0^{(0)}$, and $\cB_0^{(1)}$, we demand the equations of motion obtained from action (\ref{gravacng}) and (\ref{gravacneff}) are same at order \apt\ and order \aps\ separately. Comparing the equations of motion for $\ph$ from two actions at order \apt\ and \aps\ we get the function $\cB_1$'s and $\cB_0$'s. Explicit expression for $\cB_0$'s and $\cB_1$'s are given in appendix [\ref{cb0cb1}].

The effective coupling $K_{\rm{eff}}$ of transverse graviton is given by,
\ben
16 \pi G_5 K_{\rm{eff}}(r) &=& {(\cA^{(0)}_1(r,k) +
\ap \cB_1^{(0)}(r,k) + \ap^2 \cB_1^{(1)}(r,k))\over \sqrt{-g} g^{rr}} \nonumber \\
&=& - \frac {1} {2} + \big (20 \beta_1 +
    2 \big (r^2 - 1 \big) \beta_2 + 4
         \beta_3 \big) \ap \nonumber \\
         &&+ \frac {1} {6} \big (-36 \big (r^4 - 22 r^2 - 3 \big) \alpha_1 +
9 \big (45 r^4 + 18 r^2 - 7 \big) \alpha_2
+
   8 \big ( 3 \big (7
              r^4 - 2 r^2 - 1 \big) \alpha_3 \nn
              && -
      3 \big (9 r^4 + 10 r^2 - 5 \big)
           \alpha_4 +237 \beta_2^2 r^4 + 18 \beta_1 \beta_2 r^4
+
      66 \beta_2
           \beta_3 r^4 - 188 \beta_2^2 r^2 \nn
           && +
      10 \beta_1
           \beta_2 r^2 -
      46 \beta_2 \beta_3 r^2 + 100
           \beta_1^2 - \beta_2^2 +
      4 \beta_3^2  +
      40 \beta_1 \beta_3 \big) \big) \ap^2\ .
\een
The shear viscosity coefficient is determined by the following expression,
\ben \label{finaleta}
\eta &=& {1\over r_0^{3/2}} \lb -2 K_{\rm{eff}}(r_0) \rb \nonumber \\
&=&  \frac {1} {16  \pi {G_5}  } - \frac {(5 \beta_1 + \beta_3) \ap} {2
        \pi G_5  } \nonumber \\
        && -\ \frac {\big(108 \alpha_1 + 63 \alpha_2 +
      12 \alpha_3 - 42
           \alpha_4 + 100 \beta_1^2 +
      28 \beta_2 \beta_1 + 40
           \beta_3 \beta_1 +
      48 \beta_2^2 + 4 \beta_3^2 +
      20 \beta_2 \beta_3 \big) \ap^2} {6 \pi
        G_5 } \nonumber \\
\een
where $r_0$ is the position of horizon and in our parametrization $r_0=1$.

\subsection{Shear Viscosity to Entropy Density Ratio}

One can calculate entropy density using Wald's formula \cite{wald,dg}. Order \aps \ correction to entropy density $s$ turns out to be,
\ben \label{finals}
s &=&\frac {1} {4 {G_5}} - \frac {2 (5 \beta_1 - \beta_2 + \beta_3) \ap}
       {{G_5}} \nonumber \\
       &+&   \frac { \big (36 \alpha_1 +27 \alpha_2 -36 \alpha_4 -
      4 \big (50
              \beta_1^2 +
         4 \beta_2 \beta_1 + 20 \beta_3 \beta_1 + 26 \beta_2^2 +
         2 \beta_3^2 +
         8 \beta_2 \beta_3 \big) \big)
       \ap^2} {3 {G_5}}\ .
       \een
Then we find shear viscosity to entropy density ratio is given by,
\ben \label{etabys}
{\eta\over s} &=&  \frac {1} {4 \pi }- \frac {2
        \beta_2 \ap} {\pi } \nonumber \\
&& -
\frac {(252 \alpha_1 + 153  \alpha_2 +
      24 \alpha_3  - 120 \alpha_4 +
         56 \beta_2 (5 \beta_1 - \beta_2 + \beta_3)) \ap^2} {3 \pi } \ .
\een

Thus we see that the ratio \etabys\ depends on ambiguous coefficients $\beta_1, \beta_3, \alpha_3$ and $\alpha_4$ at order \aps. But, we will show in the next section that we can get rid of these ambiguous coefficients and express the result in terms of physical boundary parameters. To be explicit, we calculate six derivative corrections to central charges $a$ and $c$ and show that it is possible to express \etabys\ in terms of these central charges and unambiguous coefficients $\alpha_1$ and $\alpha_2$, which can be fixed by other physical boundary parameters.


\sectiono{Conformal Anomaly in Six-derivative Gravity} \label{ca}

So far we have computed shear viscosity to entropy density ratio for some gauge theory plasma whose gravity dual is governed by six derivative Lagrangian given by (\ref{6dacnreqg}).
 In this section we compute the six derivative corrections to central charges $a$ and $c$ of this dual
field theory. The holographic procedure to compute conformal anomaly form two derivative
gravity has been given in \cite{skenda} and later it has been generalized to
four derivative gravity in \cite{on,bgn}. We will follow the same
approach and carry on the analysis for six derivative terms in the action.

First we assume (can be easily checked) that the
gravity theory has a \ads\ solution even in presence of the higher derivative
terms in the action. The metric, the curvature tensors and the scalar are
given as,
\be {\label {back}}
ds^2= G^{(0)}_{\mu \nu}dx^{\mu}dx^{\nu}= {L^2 \over 4 r^2}
dr^2 + \sum_{i=1}^{d} {\eta_{ij} \over r}dx^i dx^j
\ee
and,
\be
R^{(0)}= -{d(d+1) \over L^2}, \qquad R^{(0)}_{\mu \nu}=-{d \over L^2}
G^{(0)}_{\mu \nu}, \qquad
R^{(0)}_{\mu \nu \rho \sigma}=-{1 \over L^2}(G^{(0)}_{\mu \rho}G^{(0)}_{\nu
\sigma} -G^{(0)}_{\mu \sigma}G^{(0)}_{\nu \rho}).
\ee
Here, $L$ is the corrected \ads\ radius given in (\ref{radius}) and $L=1$ when
there is no higher derivative terms present in the action. $d$ is the dimension of boundary space-time.
One can obtain the equation of motion for the action
(\ref{6dacnreqg}) following \cite{nd,jjr}. The terms in the equations of motion
containing covariant derivatives of the curvature tensors vanish for the above
background (\ref{back}). The equation finally reduces to,
\ben \label{eom}
{d(d-1) \over L^2}\,-\,12\,&=& \cu \bigg( {\beta_1\over L^4} d^2 (d+1)(d-3)+
{2 \beta_2 \over L^4}d (d-3) + {\beta_3 \over L^4}d^2(d-3) \bigg)\nn
&& - \cu^2 \bigg({4 \alpha_1 \over L^6} d(d-5)
+ {\alpha_2 \over L^6} d(d^2-6d+5)- {2
\alpha_3 \over L^6}d^2(d-5)\nn
&& \qquad +{2 \alpha_4 \over L^6} d^2(d+1)(d-5)\bigg) \ .
\een

As the \ads\ metric has a second order pole at infinity, it only induces a
conformal equivalence class $[\gz^{ij}]$ of metrics on the
boundary. Following Gauge-Gravity correspondence, the boundary field theory
effective action in large $N$ limit is,
\be
W_{FT}(\gz)= S_{grav}(g;\gz),
\ee
where $ S_{grav}(g;\gz)$ is the gravity action evaluated on classical (\ads)
configuration which approaches a representative boundary metric $\gz$. Now,
for computing conformal anomaly, we consider the following
fluctuation around (\ref{back}),
\ben {\label{fluc}}
ds^2 = G_{\mu \nu}dx^{\mu}dx^{\nu} &=& {L^2 \over 4 r^2}
dr^2 + \sum_{i=1}^{d} {g_{ij} \over r}dx^i dx^j  \qquad \rm{with}, \nn
g_{ij} &=& g_{(0)ij}+ r g_{(1)ij}+ r^2 g_{(2)ij} + r^2 (\ln r) h_{(2)ij}
+ \cdots
\quad.
\een
Here, $\gz$ is the representative boundary metric and $h_{(2)}$ is traceless
 with respect to $\gz$. The determinant of the full metric (\ref{fluc}) can be
 written as,
\ben
\sqrt{-G} &=& {L \over 2} r^{-{d \over 2}-1}\sqrt{- g_{(0)}} \bigg[1+
{r \over 2} \tr[g_{(1)}] \nn
&& +r^2 ({1 \over 2}\tr[g_{(2)}] -{1 \over 4} {\rm Tr}[(g_{(1)}^2)] \nn
&& + {1 \over 8}({\rm Tr}[g_{(1)}])^2 )\bigg]+ {\cal O}(r^3)\ .
\een

For computing the conformal anomaly of the boundary field theory, we need to
evaluate all the terms in the bulk action (\ref{6dacnreqg}) in terms of
$(\gz,\go,\gt)$. Then, we regard $\gz$ as independent field on the boundary and
solve $\go$ in terms $\gz$. As we will see, the term involving $\gt$ will
vanish on-shell (\ref{eom}). To regularize the infrared divergences of the
on-shell action, we introduce a cutoff $\epsilon$ restricting the range of $r$
integral as $r \geq \epsilon$. Then the on-shell action can be written as,
\ben \label{acexp}
S&=& S_0(\gz) \epsilon^{-{d \over 2}}+ S_1(\gz,\go) \epsilon^{-{d \over 2}-1}
\nn
&&+ \cdots \cdots + S_{\rm{ln}} \ln[\epsilon] + S_{{d \over 2}} + {\cal O}(\epsilon^{{1 \over 2}})\ .
\een
Then, the conformal anomaly ${\cal T}$ of the boundary field theory is given as,
\be
S_{\rm{ln}}=-{1 \over 2}\int d^d x \sqrt{\gz} \, {\cal T}\ .
\ee
We want to find ${\cal T}$ for $d=4$. The expressions for ${\cal T}$ for four
derivatives terms in Lagrangian are given in \cite {on,bgn}. Here we
present the computation for six-derivative terms only. The
generic structure of any term in the action has the following structure
\be \label{gen}
{1 \over 2 r^{{d\over 2}+1}}\sqrt{\gz}({\cal X}_1 + {\cal X}_2 r + {\cal X}_3 r^2 + \cdots),
\ee
where, $({\cal X}_1,{\cal X}_2,\cdots)$ are some functions of $(\gz,\go,\gt, \cdots)$. Since
we are looking for the term $S_{\rm{ln}}$ in (\ref{acexp}), we only need the terms of order ${\cal O}({1 \over r})$ in (\ref{gen}). Hence, it is enough for us to terminate
the expansion in (\ref{gen}) at ${\cal O}(r^2)$ for $d=4$. Therefore the coefficient ${\cal X}_3$ will
finally contribute to the anomaly ${\cal T}$. As we will see, this knowledge will
help us to pre-eliminate certain terms in our calculation.

We will summarize our main results for four six derivative terms in the Lagrangian (\ref{6dacnreqg}). We follow the following notations:
\bc
${\rm r}^{(0)^i}_{\ \ jkl}$ $\ra$ Riemann tensor constructed out of $\gz$.\\
${\rm r}^{(0)}_{\ \ ij}$ $\ra$ Ricci tensor constructed out of $\gz$.\\
${\rm rim}^{(0)^2} = {\rm r}^{(0)}_{\ ijkl} {\rm r}^{(0)ijkl}$, \ \ \  ${\rm ric}^{(0)^2}={\rm r}^{(0)ij} {\rm r}^{(0)}_{ij}$.\\
${\rm r}^{(0)}= \gz^{ij} {\rm r^{(0)}}_{ij}$\ .
\ec
\bi
\ii
$T_1=R^{\mu \nu}_{\rho \sigma}R^{\rho \sigma}_{\alpha \beta}
R^{\alpha \beta}_{\mu \nu}$:

Here, $(\mu, \nu)$ indices run over full five dimensional space-time. One can
split the indices in $(r; i,j)$, where $(i,j)$ runs over four dimensional
boundary space time. From the leading $r-$dependence of the curvature tensors
(appendix [\ref{apprdep}]), it is easy to see that only two combinations
$R^{ij}_{kl}R^{kl}_{mn} R^{mn}_{ij}$ and $R^{ir}_{jr} R^{jr}_{kr}
R^{kr}_{ir}$ will contribute to $S_{\rm{ln}}$. The leading $r-$dependence of other
possible combinations starts from $r^3$ and hence they do not contribute to
anomaly. The expansions of $T_1$ is\footnote{We have set $L=1$ for these
  expansion. We will put back $L$ later by dimensional analysis.},
\ben {\label {T1}}
T_1&=& R^{\mu \nu}_{\rho \sigma}R^{\rho \sigma}_{\alpha \beta}
R^{\alpha \beta}_{\mu \nu}\nn
&=& R^{ij}_{kl}R^{kl}_{mn} R^{mn}_{ij} + 8 \,R^{ir}_{jr} R^{jr}_{kr}
R^{kr}_{ir}\nn
&=& - 4 d(d+1) +12 r \bigg[{\rm r}^{(0)}+ 2 (d-1) \tr[g_{(1)}] \bigg] \nn
&&+ r^2 \bigg[-6\, {\rm rim} ^{(0)^2} -60 {\rm r}^{(0)^{ij}} g_{(1)_{ij}}+
48 (d-3) \tr[\gt] \nn
&& \qquad +12 (9 -4d)\tr[(\go)^2] -36 (\tr[\go])^2 \bigg] + {\cal O}(r^3)\ .
\een
\ii
 $T_2 = R^{\mu \nu}_{\rho \sigma}R^{\rho \tau}_{\lambda \mu}
R^{\sigma \ \lambda}_{\ \,\tau \ \nu}$:

Similarly, for $T_2$, only $R^{ij}_{kl}R^{km}_{ni} R^{l \ \ n}_{\ m \ \ j}$ and
$R^{ir}_{jr} R^{jk}_{li} R^{r \ l}_{\ \, k \ r}$ contribute to the
anomaly. The expansion is,
\ben
T_2 &=& R^{\mu \nu}_{\rho \sigma}R^{\rho \tau}_{\lambda \mu}
R^{\sigma \ \lambda}_{\ \,\tau \ \nu}\nn
&=& R^{ij}_{kl}R^{km}_{ni} R^{l \ n}_{\ \, m \ j} + 3 R^{ir}_{jr}
R^{jk}_{li} R^{r \ l}_{\ \, k \ r}\nn
&=& d (1- d^2) + 3 (d-1) r \bigg[ {\rm r}^{(0)} + 2 (d-1) \tr[\go] \bigg]\nn
&& + r^2 \bigg[ -3 \,{\rm  ric}^{(0)^2} + {3 \over 2} \,{\rm rim}^{(0)^2}
+9 (3 - d){\rm  r}^{(0)^{ij}} g_{(1)_{ij}}- 6 {\rm r}^{(0)} \tr[\go] \nn
&& \qquad  - 12 (d-1)(3-d) \tr[\gt] + (-9 d^2 + 39 d- 39) \tr[(\go)^2] \nn
&& \qquad \qquad + 3(7-4 d) (\tr[\go])^2 \bigg] + {\cal O}(r^3) \ .
\een
\ii
 $T_3=R^{\mu \nu}_{\rho \sigma}R^{\rho \sigma}_{\nu \beta}
R^{\beta}_{\mu}$:

For $T_3$, three combinations contribute. They are $R^{ij}_{kl}R^{kl}_{jm}
R^{m}_{i}$ , $R^{ri}_{rj}R^{rj}_{ir}R^{r}_{r}$ and $R^{ri}_{rj}R^{rj}_{mr}
R^{m}_{i}$. The expansion is,
\ben {\label{T3}}
T_3 &=& R^{\mu \nu}_{\rho \sigma}R^{\rho \sigma}_{\nu \beta}
R^{\beta}_{\mu}\nn
&=&  R^{ij}_{kl}R^{kl}_{jm} R^{m}_{i}+ 2 \bigg( R^{ri}_{rj}
R^{rj}_{ir}R^{r}_{r} +  R^{ri}_{rj}R^{rj}_{mr} R^{m}_{i}\bigg)\nn
&=& 2 d^2 (d+1) - 6 d r\bigg[{\rm r}^{(0)} + 2(d-1)\tr[\go] \bigg]\nn
&& + r^2 \bigg[4\, {\rm ric}^{(0)^2} + d\,{\rm rim}^{(0)^2}
+2(11 d-8) {\rm r}^{(0)^{ij}} g_{(1)_{ij}}+ 8 {\rm r}^{(0)} \tr[\go]
+ 24 d (3- d) \tr[\gt] \nn
&&\qquad + (20 d^2 -54 d+16) \tr[(\go)^2]
+ 2(11 d-8) (\tr[\go])^2 \bigg] + {\cal O}(r^3) \ .
\een
\ii
$T_4= R R^{\mu \nu}_{\rho \sigma}R^{\rho \sigma}_{\mu \nu}$:

For this term we only need to find contraction of two Riemann tensors. The
 expansion is,
\ben
T_4&=&R R^{\mu \nu}_{\rho \sigma}R^{\rho \sigma}_{\mu \nu}\nn
&=& R(R R^{ij}_{kl} R^{kl}_{ij} + 4 R R^{ir}_{jr}R^{jr}_{ir})\nn
&=&- 2 d^2(1+d)^2+ 6 r d(1+d) \bigg[ {\rm r}_{(0)}+ 2 (d-1) \tr \go  \bigg]\nn
&&+r^2\bigg[-4 {\rm r}_{(0)}^2 -d (1+d) {\rm rim}_{(0)}^2 -14 d(1+d)
{\rm r}^{(0)^{ij}} g_{(1)_{ij}}-16(d-1){\rm r}^{(0)} \tr[\go] \nn
&&\qquad + 24 d(d-3)(d+1)
\tr[\gt]
-2d (1+d)(8 d-19) \tr[(\go)^2] \nn
&&\qquad -2 (13 d^2-11 d+8)  (\tr[\go])^2 \bigg]+ {\cal O}(r^3) \ .
\een
\ei

Substituting all these expressions and the expressions for order \apt\ terms in (\ref{6dacnreqg}), we get,
\ben \label{logaction}
S_{\rm{ln}}&=& -{1 \over 2}\int dx^4 \sqrt{\gz}\bigg[\bigg(t_1\,{\rm r}^{(0)^2}
+t_2\, {\rm ric}^{(0)^2}+ t_3\, {\rm rim}^{(0)^2}\bigg) \nn
&& A\,{\rm r}^{(0)^{ij}} g_{(1)_{ij}} +B\,{\rm r}^{(0)} \tr[\go] + C\,
\tr[(\go)^2]
+D\, (\tr[\go])^2 +  E\,\tr[\gt] \bigg].
\een
where,
\ben
t_1 &=&  L \beta_1 - {4 \over L} \alpha_4, \nn
t_2 &=& L \beta_3 -{3 \over L} \alpha_2+{4 \over L} \alpha_3 \nn
t_3 &=& L \beta_2 -{6 \over L} \alpha_1+{3 \over 2L}
\alpha_2+{4 \over L} \alpha_3 -{20 \over L} \alpha_4
\een
and
\ben
A&=&-\frac{L}{16 G_5 \pi }
+ \cu \left(\frac{5 \beta_1}{2 G_5 L \pi }+\frac{3
   \beta_2}{4 G_5 L \pi }+\frac{3 \beta_3}{4
    G_5 L \pi }\right)\nn
&&- \cu^2 \bigg(\frac{15 \alpha_1 }{4 G_5 L^3 \pi }+\frac{9
   \alpha_2 }{16 G_5 L^3 \pi }-\frac{9 \alpha_3
   }{2 G_5 L^3 \pi }+\frac{35 \alpha_4}{2 G_5
   L^3 \pi }\bigg)\nn
B &=&\frac{L}{32 G_5 \pi }
- \cu \left(\frac{\beta_1}{2 G_5 L \pi }+\frac{
   \beta_2}{8 G_5 L \pi }+\frac{ \beta_3}{8
    G_5 L \pi }\right)\nn
&&+ \cu^2 \bigg(\frac{3 \alpha_1 }{8 G_5 L^3 \pi }-\frac{3
   \alpha_2 }{32 G_5 L^3 \pi }-\frac{\alpha_3
   }{4 G_5 L^3 \pi }+\frac{3 \alpha_4}{4 G_5
   L^3 \pi }\bigg)\nn
C &=&\frac{1}{8 G_5 \pi L}-{3 L \over 16 G_5 \pi}
+ \cu \left(\frac{5 \beta_1}{4 G_5 L^3 \pi }+\frac{5
   \beta_2}{8 G_5 L^3 \pi }+\frac{ \beta_3}{2
    G_5 L^3 \pi }\right)\nn
&&- \cu^2 \bigg(\frac{4 \alpha_1 }{ G_5 L^5 \pi }+\frac{3
   \alpha_2 }{4 G_5 L^5 \pi }-\frac{5 \alpha_3
   }{ G_5 L^5 \pi }+\frac{20 \alpha_4}{ G_5
   L^5 \pi }\bigg)\nn
D &=&-\frac{1}{32 G_5 \pi L}+{3 L \over 32 G_5 \pi}
+ \cu \left(\frac{3 \beta_1}{8 G_5 L^3 \pi }+\frac{
   \beta_2}{16 G_5 L^3 \pi }+\frac{ \beta_3}{8
    G_5 L^3 \pi }\right)\nn
&&- \cu^2 \bigg(\frac{5 \alpha_1 }{8 G_5 L^5 \pi }+\frac{15
   \alpha_2 }{32 G_5 L^5 \pi }-\frac{5 \alpha_3
   }{4 G_5 L^5 \pi }+\frac{21 \alpha_4}{4 G_5
   L^5 \pi }\bigg) \nn
E &=&{1 \over 16 G_5 \pi}\bigg[-{6 \over L} +6 L +\cu \bigg( {40 \beta_1
\over L^3}+ {4 \beta_2 \over L^3}+{8 \beta_3 \over L^3} \bigg) \nn
&& + \cu^2 \bigg({8 \alpha_1 \over L^5}+{6 \alpha_2 \over L^5}-{16 \alpha_3
 \over L^5}+{80 \alpha_4 \over L^5}\bigg) \bigg]\ .
\een
It is easy to see that $\tr[\gt]$ term vanishes when the equation of
motion (\ref{eom}) is satisfied. The equation and the solution for $\go$ are given by
\ben
A r_{(0)}^{ij}+ B \gz^{ij}r_{(0)} r+2 C \gz^{ik}\gz^{jl}g_{(1)kl}
+ 2D\gz^{ij}\gz^{kl}g_{(1)kl}=0
\een
and
\ben
g_{(1)ij}=-{A \over 2C}r_{(0)ij}+{A D -B C \over 2 C (C+4D)} r_{(0)}
g_{(0)ij}.
\een
We can also rearrange equation (\ref{eom}) to write the corrected \ads\ radius as (for $d=4$),
\ben \label{radius}
L &=& 1- {1 \over 3}\cu(10 \beta_1-\beta_2-2 \beta_3)
+{1 \over 18}\cu^2
(-12 \alpha_1-9 \alpha_2+24 \alpha_3 -120 \alpha_4\nn
&&\qquad -500 \beta_1^2-5 \beta_2^2
-20 \beta_3^2 -100
\beta_1 \beta_2-200 \beta_1 \beta_3-20 \beta_2 \beta_3)\ .
\een

Substituting all these expression in (\ref{logaction}), we get the conformal
anomaly as,
\ben
{\cal T}&=& - a E_4 -c I_4 \nn
&=& - a (r_{(0)}^2 -4 {\rm ric}_{(0)}^2 + {\rm rim}_{(0)}^2)+ c ({1 \over 3}
 r_{(0)}^2 - 2 {\rm ric}_{(0)}^2 + {\rm rim}_{(0)}^2),
\een
 where the coefficients $a$ and $c$ are given as,
\ben
a &=&\frac{1}{128 G_5 \pi }-\cu \frac{5 (10 \beta_1+\beta_2 +2 \beta_3)}{128
(G_5 \pi )}\nn
&&+ \cu^2 \frac{60 \alpha_1 + 45 \alpha_2 +5 \left((10
\beta_1+ \beta_2+2 \beta_3)^2-24
\alpha_3+120
   \alpha_4\right)}{768 G_5 \pi
   }
\een
and
\ben
c&=&\frac{1}{128 G_5 \pi }-\cu \frac{(50 \beta_1-3 \beta_2+10 \beta_3)}{128 G_5
 \pi }\nn
&&+ \cu^2 \frac{\left(500
   \beta_1^2-60 \beta_2 \beta_1+200
   \beta_3 \beta_1-11 \beta_2^2+20
   \beta_3^2 -12 \beta_2
   \beta_3\right)}{768 G_5 \pi }\nn
&&- \cu^2 \frac{(228 \alpha_1-117 \alpha_2-72
   \alpha_3+360 \alpha_4) }{768 G_5 \pi }\ .
\een

\sectiono{$\eta$, $s$ and \etabys} \label{etasandetabys}

It is interesting to compute the following combination,
\be \label{c-a}
{c-a \over c}=8 \cu \beta_2 +\frac{4}{3}\cu^2 (-36 \alpha_1
+9 \alpha_2+4 (6 \alpha_3-30 \alpha_4+\beta_2 (70
   \beta_1-5 \beta_2+14 \beta_3)))\ .
\ee
 From the above relation (\ref{c-a}) and (\ref{temp}), (\ref{finaleta}), (\ref{finals}) and (\ref{etabys}), one can see that the the
 ambiguous coefficients $(\beta_1, \beta_3, \alpha_3, \alpha_4)$ appear in
 $s$, $\eta$ and  ${\eta \over s}$ and ${c-a \over c}$ in such a way that one can replace them in terms of this combination of central charges. Hence, we can
 rewrite $\eta$, $s$ and ${\eta \over s}$ as,
\be
\eta= 8 \pi^3 c \ T^3 \bigg[ 1 + {1\over 4}{c-a \over c} - {1\over 8} \bigg( {c-a \over c}\bigg)^2- 180 \cu^2 (2 \alpha_1 +  \alpha_2)\bigg]  + {\cal O}(\alpha'^3)\ ,
\ee
 \be
s = 32 \pi^4 c\  T^3 \bigg[ 1 + {5\over 4} {c-a \over c}  +
 {3 \over
8}\bigg({c-a \over c}\bigg)^2+ 12 \cu^2(2 \alpha_1+ \alpha_2)\bigg]  + {\cal O}(\alpha'^3)
\ee
 and
\be
{\eta \over s} = {1\over 4 \pi}\bigg[1 -{c-a \over  c} + {3\over 4} \bigg({c-a \over c}\bigg)^2 - 192  \alpha'^2 (2 \alpha_1 + \alpha_2)\bigg] + {\cal O}(\alpha'^3)\ .
\ee

These are the main results of this paper. Here, we have been able to
rewrite shear viscosity $\eta$, entropy density $s$ and the ratio \etabys\ in terms of central charges $c$ and $a$ of
boundary field theory and two other unambiguous parameters $\alpha_1$
and $\alpha_2$.

In \cite{hofmal} the authors considered energy correlation function which is quantum expectation value of a product of energy flux operators on the state produced by the localized operator insertion,
\be
\langle {\cal E}(\theta_1) \cdots {\cal E}(\theta_n)\rangle \equiv {\langle 0| {\cal O}^{\dag} {\cal E}(\theta_1) \cdots {\cal E}(\theta_n) {\cal O} | 0\rangle \over \langle 0| {\cal O}^{\dag} {\cal O} | 0\rangle }
\ee
where ${\cal O}$ is the operator creating the localized state and $\theta_1 \cdots \theta_n$ are the (angular)positions of the calorimeters which measures the total energy per unit angle deposited at each of these angles. In particular they considered energy one point function $\langle {\cal E}(\theta)\rangle$ when states are created by stress tensor. This energy one point function is basically three point correlation function of CFT stress tensors. The most general expression for this energy one point function is
\be
\langle {\cal E}(\theta)\rangle = {\langle 0| \epsilon^{*}_{ij} T_{ij} {\cal E}(\theta) \epsilon_{lk} T_{lk}|0\rangle \over  \langle 0| \epsilon^{*}_{ij} T_{ij}\epsilon_{lk} T_{lk}|0\rangle}
={q^0\over 4 \pi} \bigg[1 + \tau_2 \bigg({\epsilon^{*}_{ij} \epsilon_{il}n_jn_l \over \epsilon^{*}_{ij}\epsilon_{ij}} -{1\over 3} \bigg) + \tau_4 \bigg({|\epsilon_{ij}n_in_j|^2 \over \epsilon^{*}_{ij}\epsilon_{ij}} - {2\over 15}\bigg)\bigg]
\ee
where $\epsilon_{ij}$ is symmetric polarization tensor and $\theta$ is the angle between the point on $S^2$, labeled by $n_i$.

There are two undetermined parameters $\tau_2$ and $\tau_4$. In \cite{hofmal}, it has been shown that these two parameters can be related to the coefficients multiply higher order gravity correction. When the dual gravity theory is governed by Einstein-Hilbert action (no higher derivative terms) then these two parameters turn out to be zero. In higher derivative bosonic theory when one considers terms like
$$
\ap \beta_2 R_{\mu\nu\rho\sigma}R^{\mu\nu\rho\sigma} + \ap^2 \bigg[ \alpha_1 I_1 + \alpha_2 I_2\bigg] $$
then these two parameters are related to the coefficients of higher derivative terms,
$$
\tau_2 \sim \ap \beta_2 + {\cal O}(\ap^2) \qquad \qquad {\rm {and}} \qquad \qquad
\tau_4 \sim \ap^2\ f(\alpha_1,\alpha_2)\ ,
$$
where, $f$ are some linear functions in $\alpha_1$ and $\alpha_2$ ($\sim 2 \alpha_1 + \alpha_2$).
$\tau_2$ is also related to central charges $a$ and $c$ of the theory ($\tau_2 \sim (c-a)/c$). Hence $\beta_2$ is fixed in terms of central charges (at order \apt) \cite{on,bgn}
 and $f$ is fixed in terms of $\tau_4$ at order \aps. Since all physical quantities depend on a particular combination  $2\alpha_1 + \alpha_2$ of unambiguous coefficients therefore we can completely fix them in terms of CFT parameters $c$, $a$ and $\tau_4$.

Thus we see that the physical measurable quantities $\eta$, $s$ and  \etabys\
of boundary field theory are finally independent of
ambiguous parameters and completely depend on physical boundary parameters.


\bc
{\bf ---------------------------- }
\ec

\vspace{1.5cm}
\bc
{\bf Acknowledgement}
\ec
We are thankful to Justin David, Rajesh Gopakumar, Dileep Jatkar, Ayan
Mukhopadhyay and Ashoke Sen for useful discussions. We are specially
grateful to Rajesh Gopakumar for his valuable comments on the
manuscript. We are also thankful to Robert Myers for his critical
comments on the first version of our
paper. SD would like to acknowledge NSERC of Canada. Finally we would
like to thank people of India for their kind support to fundamental
research.


\newpage
\noindent
{\bf \Large Appendix}

\appendix

\sectiono{Expressions for $A_i$'s}\label{As}

Expressions for $A_i$'s in $k\ra 0$ limit are given by,
\ben
A_1(r) &=& - {1 \over 9 r^3}(\big (2592 \beta_2^2 r^6 +
      6048 \alpha_3 r^6 - 27648 \alpha_4 r^6 +
      18432 \beta_1 \beta_2 r^6 \nonumber \\
      && +
      5760 \beta_2
           \beta_3 r^6 - 4788 \beta_2^2 r^4 -
      3168 \alpha_3 r^4 + 13392 \alpha_4 r^4 -
      9288 \beta_1 \beta_2 r^4 - 3816
           \beta_2 \beta_3 r^4 \nonumber \\
           && +
      6500 \beta_1^2 + 65 \beta_2^2 +
      260 \beta_3^2 -
      12 \big (816 r^6 - 558 r^4 + 19 \big) \alpha_1 -
      \big (2880 r^6 - 1134 r^4 - 171 \big) \alpha_2 \nonumber \\
           && + 456 \alpha_3 -
      2280 \alpha_4 +
      1300 \beta_1 \beta_2 +
      2600 \beta_1
           \beta_3 +
      260 \beta_2 \beta_3 \big) \ap^2) \nonumber \\
      && - \frac {\big (36 \beta_2 r^4 +
      220 \beta_1 + 22 \beta_2 + 44
           \beta_3 \big) \ap} {3 r^3} + \frac {2} {r^3}\nn
A_2(r) &=& {1\over 6 r}(\big (2736 \beta_2^2 r^6 +
      1152 \alpha_3 r^6 - 3456 \alpha_4 r^6 +
      2304 \beta_1 \beta_2 r^6 +
      1248 \beta_2
           \beta_3 r^6 - 6996 \beta_2^2 r^4 \nonumber \\
           && +
      48 \beta_3^2 r^4 - 432 \alpha_3 r^4 +
      1104 \alpha_4 r^4 -
      1416 \beta_1 \beta_2 r^4 + 240 \beta_1 \beta_3 r^4 -
      2448 \beta_2 \beta_3 r^4 \nonumber \\
      &&+
      6500 \beta_1^2 r^2 + 3701 \beta_2^2 r^2
      356
           \beta_3^2 r^2 + 984 \alpha_3 r^2 -
      2808 \alpha_4 r^2 +
      1852 \beta_1 \beta_2 r^2 \nonumber \\
      &&+
      3080 \beta_1 \beta_3
          r^2 + 1436 \beta_2 \beta_3 r^2 -
      6500 \beta_1^2 - 17 \beta_2^2 -
      212 \beta_3^2 -
      12 \big (312 r^6 \nonumber \\
      &&- 318 r^4 + 193 r^2 + 5 \big)
           \alpha_1 +
      9 \big (24 r^6 + 58 r^4 - 69 r^2 + 19 \big) \alpha_2 - 168 \alpha_3 \nonumber \\
      &&+
      1320 \alpha_4 -
      820 \beta_1 \beta_2 - 2360
           \beta_1 \beta_3 -
      140 \beta_2 \beta_3 \big) \ap^2)   \nonumber \\
      &&+ \frac {\big (110 \big (r^2 - 1 \big) \beta_1 + \big (34
              r^4 + r^2 - 3 \big) \beta_2 +
      2 \big (2 r^4 + 15 r^2 - 9 \big) \beta_3 \big) \ap} {r} - 3 r + \frac {3} {r}
\nn
A_3(r)&=& \ap r \big (r^2 -
     1 \big)^2 (4 \beta_2 + \beta_3) -
 4 \ap^2 r \big (r^2 - 1 \big)^2 \big (16 \beta_2^2 r^2 - 4 \alpha_3 r^2 \nonumber \\
 && +
    4 \beta_2 \beta_3 r^2 - 4 \beta_2^2 -
    2 \beta_3^2 +
    24 \big (r^2 + 1 \big) \alpha_1 +
    3 \big (r^2 - 1 \big) \alpha_2 - 20 \alpha_3 + 80 \alpha_4 \nonumber \\
 && -
    40 \beta_1 \beta_2 -
    10 \beta_1 \beta_3 -
    9 \beta_2 \beta_3 \big)
\nn
A_4(r) &=&  {1 \over 9 r^2}(\big (-72 \big (372 \alpha_1 -
         2 \big (241 \beta_2^2 +
            96 \beta_1 \beta_2 +
            100 \beta_3
                 \beta_2 + 49 \alpha_3 -
            144 \alpha_4 \big) \big) r^6 \nonumber \\
            &&+ \big( 2106 \alpha_2 +
      36 \big (-1049 \beta_2^2 -
         138 \beta_1 \beta_2 - 338
              \beta_3 \beta_2 +
         642 \alpha_1  -
         104 \alpha_3 +
         132 \alpha_4 \big)\big) r^4 \nonumber \\
            &&+ \big (6500 \beta_1^2 + 1372 \beta_2 \beta_1 +
         2600 \beta_3 \beta_1 +
         3557 \beta_2^2 + 260 \beta_3^2 -
         2028 \alpha_1 - 477
              \alpha_2 + 600 \alpha_3 \nonumber \\
            &&-
         1848 \alpha_4 +
         1196 \beta_2 \beta_3 \big) r^2 +
      19500 \beta_1^2 + 195 \beta_2^2 + 780 \beta_3^2 -
      684 \alpha_1 - 513 \alpha_2 +
      1368 \alpha_3 \nonumber \\
            &&- 6840 \alpha_4 +
      3900 \beta_1 \beta_2 + 7800
           \beta_1 \beta_3 +
      780 \beta_2 \beta_3 \big) \ap^2) \nonumber \\
            &&+ \frac {2 \big (110 \big (r^2 +
          3 \big) \beta_1 + \big (90 r^4 - 7
              r^2 + 33 \big) \beta_2 +
      22 \big (r^2 + 3 \big) \beta_3 \big) \ap} {3 r^2} - \frac {6} {r^2} - 2
\nn
A_5(r)&=& {2 \big (r^2 -
       1 \big)\over 9 r} \big (3600 \beta_2^2 r^4 +
      1008 \alpha_3 r^4 - 3456 \alpha_4 r^4 +
      2304 \beta_1 \beta_2 r^4 + 1728
           \beta_2 \beta_3 r^4 \nonumber \\
            &&-
      3492 \beta_2^2 r^2 - 144 \alpha_3 r^2 -
      432 \alpha_4 r^2 -
      72 \beta_1 \beta_2 r^2 - 936
           \beta_2 \beta_3 r^2 +
      6500 \beta_1^2 + 65 \beta_2^2 \nonumber \\
            &&+
      260 \beta_3^2 -
      12 \big (264 r^4 - 150 r^2 + 19 \big) {\alpha_1} + 9 \big (16 r^4 + 34 r^2 - 19 \big) \alpha_2 + 456 \alpha_3 \nonumber \\
            &&-
      2280 \alpha_4 +
      1300 \beta_1 \beta_2 +
      2600 \beta_1
           \beta_3 +
      260 \beta_2 \beta_3 ) \ap^2 \nonumber \\
            &&+ \frac {4 \big (r^2 -
       1 \big) \big (110 \beta_1 + \big (18 r^2 + 11 \big) \beta_2 + 22 \beta_3 \big) \ap} {3 r} -
 4 r + \frac {4} {r}
\nn
A_6(r)&=& 8 \big (r^2 -
     1 \big) \big (-\big (24 \beta_2^2 +
       8 \beta_3 \beta_2 +
       24 \alpha_1 -3 \alpha_2 -
       6 \alpha_3 \big)
        r^4 + \big (12 \beta_2^2 +
       40 \beta_1 \beta_2 \nonumber \\
            &&+
       9 \beta_3 \beta_2 +
       2 \beta_3^2 - 24 \alpha_1 + 6
            \alpha_2 + 22 \alpha_3 -
       80 \alpha_4 \nonumber \\
            &&+
       10 \beta_1 \beta_3 \big) r^2 +
    2 \beta_3^2 - 3 \alpha_2 + 4
         \alpha_3 +
    10 \beta_1 \beta_3 + \beta_2 \beta_3 \big) \ap^2 \nonumber \\
            &&+
 8 \big (r^2 -
     1 \big) \big ((4 \beta_2 + \beta_3) r^2 + \beta_3 \big) \ap\ .
\een

\sectiono{Expressions for $\cB_0$ and $\cB_1$}\label{cb0cb1}

\ben
\cB_0^{(0)} &=& \frac {\big (130 {\beta_1} \omega^2
+ 6 r^2 \beta_2 \omega^2 -
      11 \beta_2
           \omega^2 +
      26 \beta_3 \omega^2 \big) } {12 r^2
        \big (r^2 -
      1 \big)}  \nonumber \\
      \cB_0^{(1)}      &=&  {1\over 72 r^2 \big (r^2 -
      1 \big)}(\big (-4248 \beta_2^2 \omega^2 r^4 +
      12240 \alpha_1
           \omega^2 r^4 + 5148 \alpha_2 \omega^2 r^4 -
      432 \alpha_3 \omega^2 r^4 \nonumber \\
      && +
      864 \alpha_4 \omega^2 r^4 - 576 \beta_1 \beta_2 \omega^2 r^4 -
      1152 \beta_2 \beta_3 \omega^2 r^4 +
      5916 \beta_2^2 \omega^2 r^2 - 6984
           \alpha_1 \omega^2 r^2 \nonumber \\
      && -
      1818 \alpha_2 \omega^2 r^2 + 864
           \alpha_3 \omega^2 r^2 -
      2448 \alpha_4 \omega^2
          r^2 +
      1272 \beta_1 \beta_2 \omega^2 r^2 + 1752
           \beta_2 \beta_3 \omega^2 r^2 \nonumber \\
      &&+
      2900 \beta_1^2
           \omega^2 - 19 \beta_2^2 \omega^2 +
      116 \beta_3^2 \omega^2 + 660 \alpha_1 \omega^2 -
      369 \alpha_2 \omega^2 - 168
           \alpha_3 \omega^2 +
      840 \alpha_4 \omega^2 \nonumber \\
      &&+ 100
           \beta_1 \beta_2 \omega^2 +
      1160 \beta_1
           \beta_3 \omega^2 +
      20 \beta_2 \beta_3 \omega^2 \big))
\een

\ben
\cB_1^{(0)} &=& - \frac {\big (r^2 -
       1 \big) \big (18 \beta_2 r^2 + 110 \beta_1 - 13 \beta_2 +
      22 \beta_3 \big) } {3 r} \nonumber \\
\cB_1^{(1)} &=& {1\over 18 r}(\big (r^2 -
       1 \big) \big (-15408 \beta_2^2 r^4 + 3168
           \alpha_1 r^4 - 2304 \alpha_2 r^4 -
      1728 \alpha_3
          r^4 + 3456 \alpha_4 r^4 \nonumber \\
&&-
      2304 \beta_1 \beta_2
          r^4 - 4608 \beta_2 \beta_3 r^4 +
      12420 \beta_2^2
          r^2 - 6984 \alpha_1 r^2 -
      1170 \alpha_2 r^2 + 720
           \alpha_3 r^2\nonumber \\
&& + 432 \alpha_4 r^2 +
      72 \beta_1
           \beta_2 r^2 +
      3240 \beta_2 \beta_3 r^2 - 6500
           \beta_1^2 + 79 \beta_2^2 -
      260 \beta_3^2 - 636
           \alpha_1 + 387 \alpha_2 \nonumber \\
&&+
      120 \alpha_3 - 600
           \alpha_4 +
      140 \beta_1 \beta_2 - 2600
           \beta_1 \beta_3 +
      28 \beta_2 \beta_3 \big) \ap) \ .
\een


\sectiono{Shear Viscosity from Kubo's Formula}\label{kubo}

The shear viscosity coefficient of boundary fluid is related to the imaginary
part of retarded Green function in low frequency limit. The retarded
Green function $G^R_{xy,xy}(k)$
is defined in the following way. The on-shell action for
graviton can
be written as a surface term,
\ben \label{rgf}
S &=&   \nt \intk\,\phi_0(k)\,{\cal G}_{xy,xy}(k,r)\,\phi_0(-k)\bigg|_{r=0}\nn
&=& \nt \intk\, {\cal F}_{xy,xy}(k)
\een
where $\phi_0(k)$ is the boundary value of $\ph$
and $G^{R}_{xy,xy}$ is given by,
\ben
G^R_{xy,xy}(k)=\lim_{r\ra 0}2 {\cal G}_{xy,xy}(k,r)
\een
and shear viscosity coefficient is given by,
\ben \label{etadef}
\eta = \lim_{\omega \ra 0} \bigg[ {1\over \omega}{\rm Im} G^R_{xy,xy}(k)
  \bigg ] \ .
\een
To calculate this number
  one has to know the exact solution, \ie the form of $\ph$. The solution for $\ph$ up to order \aps\ is given by,
\ben
\phi(r,k)&=& 1 -  \mathfrak{i} \beta  \omega  \log \big (1 - r^2 \big) -6 \mathfrak{i} \cu \beta \beta_2 \omega r^2 +
 2 \mathfrak{i} \cu^2 \beta  \big (-223
          \beta_2^2 r^2 - 24 \alpha_3 r^2 +
     48 \alpha_4
         r^2 \nonumber \\
         && - 32 \beta_1 \beta_2 r^2 -
     64 \beta_2
          \beta_3 r^2 - 70 \beta_2^2 +
     2 \big (22 r^2 - 53 \big )
          \alpha_1 - \big (32 r^2 + {193 \over 2} \big) \alpha_2 \nonumber \\
         && - 28
          \alpha_3 + 108 \alpha_4 -
     172 \beta_1
          \beta_2 -
     60 \beta_2 \beta_3 \big) \omega
   r^2
   \een
where,
\ben
\beta&=& \sqrt{-{g_{rr}\over g_{tt}} (1-r)^2}\ .
\een
With this solution we calculate ${\cal F}_{xy,xy}(k)$ after adding proper $Gibbons-Hawking$ boundary terms to the action (\ref{gravacn}). Then we find shear viscosity coefficient $\eta$ from imaginary part of ${\cal F}_{xy,xy}(k)$ following equation (\ref{etadef}). It turns out that,
\ben
\eta &=& \frac {1} {16  \pi {G_5}  } - \frac {(5 \beta_1 + \beta_3) \ap} {2
        \pi G_5  } \nonumber \\
        && -\ \frac {\big(108 \alpha_1 + 63 \alpha_2 +
      12 \alpha_3 - 42
           \alpha_4 + 100 \beta_1^2 +
      28 \beta_2 \beta_1 + 40
           \beta_3 \beta_1 +
      48 \beta_2^2 + 4 \beta_3^2 +
      20 \beta_2 \beta_3 \big) \ap^2} {6 \pi
        G_5 } \ . \nonumber \\
\een


\sectiono{Leading $r-$Dependence of Curvature Tensors}\label{apprdep}

In this appendix, we give the $r-$dependence of various Riemann and Ricci
tensors. As discussed in section (\ref{ca}) below equation (\ref{gen}), while
computing the four and six derivative terms, we need to keep those terms up to
order $r^2$. If for some combinations, the leading $r-$dependence starts from
order $r^3$, they will not contribute to anomaly.
\ben \label{rijkl}
R_{ijkl} &=& r^{-2}[g_{(0)_{il}}g_{(0)_{jk}}-g_{(0)_{ik}}g_{(0)_{jl}}]\nn
         &+& r^{-1} r^{(0)}_{ijkl}\nn
         &+&  [ g_{(0)_{ik}}\ff_{jl} +g_{(0)_{jl}}\ff_{ik}
                       -g_{(0)_{il}}\ff_{jk} -g_{(0)_{jk}}\ff_{il}]\nn
         &+&  [\nabla^{0}_{k}\delta\Gamma_{ijl}-\nabla^{(0)}_{l} \delta\Gamma_{ijk}]
\nn
         &+&  [g_{(2)_{in}}r^{(0)n}_{\;jkl}] + {\cal O}(r)\ .
\een
\ben
R_{rijk} &=&
r^{-1}[-\frac{1}{2}(\nabla_{j} g_{(2)_{ik}}-\nabla_{k} g{(2)_{ij}})]
+ {\cal O}(1)\ .
\een
\ben
R^{r}_{\;i r j}&=& r ^{-1}[- g{(0)_{ij}}]\nn
                    &+& r^{0} [-g_{(0)_{ij}}]\nn
                    &+& r^{+1}[-5 \ff_{ij} + (g_{(2)})^{2}_{ij}]+ {\cal
O}(r^{2})\;\;,
\label{rij}
\een
\ben
R^{ij}_{\ \ kl} = {\cal O}(1) \qquad && \qquad
R^{ir}_{\ \ kr} = {\cal O}(1)\nn
R^{ir}_{\ \ kl} = {\cal O}(r^2)\qquad && \qquad
R^{ij}_{\ \ kr} = {\cal O}(r)\nn
R^{r}_r = {\cal O}(1)\qquad && \qquad
R^{i}_{r} = {\cal O}(r)\nn
R^{r}_{i} = {\cal O}(r^2)\qquad && \qquad
R^{i}_{j} = {\cal O}(1)\ .
\een


\end{document}